\documentclass[reprint,superscriptaddress,secnumarabic,amssymb, nobibnotes aps, prd]{revtex4-1}
\usepackage{braket}
\usepackage{graphicx}
\usepackage{amsmath}
\usepackage{mathtools}
\providecommand{\abs}[1]{\lvert#1\rvert}
\providecommand{\hdag}[2]{\hat{a}_#1^{\dagger #2}}
\begin{document}

\title{Inefficiency of classically simulating linear optical quantum computing with Fock-state
inputs}
\author{Bryan T. Gard}
\email{bgard1@lsu.edu}
\affiliation{Department of Physics \& Astronomy, Hearne Institute for Theoretical Physics, Louisiana State University, 202 Nicholson Hall, Baton Rouge, Louisiana 70803, USA}
\author{Jonathan P. Olson}
\affiliation{Department of Physics \& Astronomy, Hearne Institute for Theoretical Physics, Louisiana State University, 202 Nicholson Hall, Baton Rouge, Louisiana 70803, USA}
\author{Robert M. Cross}
\affiliation{Department of Physics \& Astronomy, University of Rochester, P.O. Box 270171, Rochester, New York 14627, USA}
\author{Moochan B. Kim}
\affiliation{Department of Physics \& Astronomy, Hearne Institute for Theoretical Physics, Louisiana State University, 202 Nicholson Hall, Baton Rouge, Louisiana 70803, USA}
\author{Hwang Lee}
\affiliation{Department of Physics \& Astronomy, Hearne Institute for Theoretical Physics, Louisiana State University, 202 Nicholson Hall, Baton Rouge, Louisiana 70803, USA}
\author{Jonathan P. Dowling}
\affiliation{Department of Physics \& Astronomy, Hearne Institute for Theoretical Physics, Louisiana State University, 202 Nicholson Hall, Baton Rouge, Louisiana 70803, USA}
\affiliation{Computational Science Research Center, No.3 HeQing Road, Beijing, China}

\date{\today}
\begin{abstract}
Aaronson and Arkhipov recently used computational complexity theory to argue that classical computers very likely cannot efficiently simulate linear, multimode, quantum\textendash optical interferometers with arbitrary Fock\textendash state inputs [S. Aaronson and A. Arkhipov, Theory Comput., \textbf{9}, 143 (2013)]. Here we present an elementary argument that utilizes only techniques from quantum optics. We explicitly construct the Hilbert space for such an interferometer and show that its dimension scales exponentially with all the physical resources. We also show in a simple example just how the Schr\"{o}dinger and Heisenberg pictures of quantum theory, while mathematically equivalent, are not in general computationally equivalent.  Finally, we conclude our argument by comparing the symmetry requirements of multi\textendash particle bosonic to fermionic interferometers and, using simple physical reasoning, connect the non\textendash simulatability of the bosonic device to the complexity of computing the permanent of a large matrix. 

\end{abstract}
\maketitle

\section{Introduction}
There is a history of attempts to use linear quantum interferometers to design a quantum computer. \v{C}ern\'{y} showed that a linear interferometer could solve NP\textendash complete problems in polynomial time but only with an exponential overhead in energy \cite{cerny}. Clauser and Dowling showed that a linear interferometer could factor large numbers in polynomial time but only with exponential overhead in both energy and spatial dimension \cite{clauser}. Cerf, Adami, and Kwiat showed how to build a programmable linear quantum optical computer but with an exponential overhead in spatial dimension \cite{cerf}.

Nonlinear optics provides a well\textendash known route to universal quantum computing \cite{milburn}. We include in this nonlinear class the so\textendash called \lq\lq{}linear\rq\rq{} optical approach to quantum computing \cite{knill}, because this scheme contains an effective Kerr nonlinearity \cite{lapaire}.

In light of these results there arose a widely held belief that linear interferometers alone, even with nonclassical input states, cannot provide a road to universal quantum computation and, as a corollary, that all such devices can be efficiently simulated classically. However, recently Aaronson and Arkhipov (AA) gave an argument that multimode, linear, quantum optical interferometers with arbitrary Fock\textendash state photon inputs very likely could not be simulated efficiently with a classical computer \cite{aaronson}. Their argument, couched in the language of quantum computer complexity class theory, is not easy to follow for those not skilled in that art. Nevertheless, White and collaborators, and several other groups, carried out experiments that demonstrated that the conclusion of AA holds up for small photon numbers \cite{broome,crespi,tillmann,spring}. Our goal here is to understand\textemdash from a physical point of view\textemdash why such a device cannot be simulated classically.

\begin{figure}
\centering
\includegraphics[height=4.5cm]{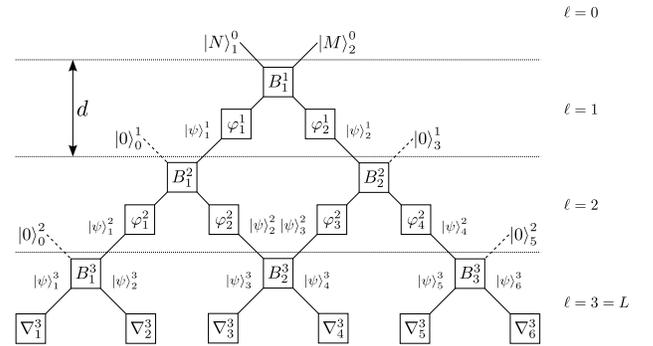}
\caption{Quantum pachinko machine for numerical depth $L = 3$. We indicate an arbitrary bosonic dual\textendash Fock input $\ket{N}\ket{M}$ at the top of the interferometer and then the lattice of beam splitters ($B$), phase shifters ($\varphi$), and photon\textendash number\textendash resolving detectors ($\nabla$). The vacuum input modes $\ket{0}$  (dashed lines) and internal modes $\ket{\psi}$ (solid lines) are also shown. The notation is such that the superscripts label the level $\ell$ and the subscripts label the row element from left to right.}
\label{fig:fig1}
\end{figure}

In their paper, AA prove that both strong and weak simulation of such an interferometer is not efficient classically.  In the context of Fock-state interferometers, a strong simulation implies the direct computation of the joint output probabilities of a system.  However, one can consider a \lq\lq{}weak\rq\rq{} simulation where one could efficiently estimate the joint output probabilities to within some acceptably small margin of error.  There are many examples of systems for which weak simulation is efficient even when strong simulation is not, such as finding the permanent of an $n\times n$ matrix with real, positive entries. But as our goal is to provide the most straightforward and physical explanation for this phenomenon, we do so only for the strong case.  Since many classical systems cannot even be strong simulated, it may at first seem unsurprising that this is the case.  However, we note that here not only does our system's classical counterpart\textemdash Galton's board\textemdash admit an efficient strong simulation, but so does a myriad of other quantum interferometers with non-Fock state inputs as we will show.

We then independently came to the same conclusion as AA in our recent analysis of multi\textendash photon quantum random walks in a particular multi\textendash mode interferometer called a quantum \lq\lq{}pachinko\rq\rq{} machine shown in Fig.\ref{fig:fig1} \cite{gard}. The dual\textendash photon Fock state $\ket{N}\ket{M}$ is inputted into the top of the interferometer and then the photons are allowed to cascade downwards through the lattice of beam splitters ($B$) and phase shifters ($\varphi$) to arrive at an array of photon\textendash number\textendash resolving detectors ($\nabla$) at the bottom. Our goal was to compute all the joint probabilities for, say, the $q^{th}$ detector received $p$ photons while the $r^{th}$ detector received $s$ photons, and so forth, for arbitrary input photon number and lattice dimension. We failed utterly. It is easy to see why.

Working in the  Schr\"{o}dinger picture, we set out to compute the probability amplitudes at each detector by following the Feynman\textendash like paths of each photon through the machine, and then summing their contributions at the end. For a machine of numerical depth $L$, as shown in Fig. \ref{fig:fig1}, it is easy to compute that the number of such Feynman\textendash like paths is $2^{L(N+M)}$. So for even a meager number of photons and levels the solution to the problem by this Schr\"{o}dinger picture approach becomes rapidly intractable. For example, choosing $N= M = 9 $ and $L = 6$, we have $2^{288} \cong 5 \times 10^{86}$ total possible paths, which is about four orders of magnitude larger than the number of atoms in the observable universe. We were puzzled by this conclusion; we expected any passive linear quantum optical interferometer to be efficiently simulatable classically. With the AA result now in hand, we set out here to investigate the issue of the complexity of our quantum pachinko machine from an intuitive physical perspective. The most mathematics and physics we shall need is elementary combinatorics and quantum optics.

Following Feynman, we shall explicitly construct the pachinko machine's Hilbert state space for an arbitrary level $L$, and for arbitrary photon input number, and show that the space's dimension grows exponentially as a function of each of the physical resources needed to build and run the interferometer \cite{feynman}.  Because interference only occurs when the input state has been symmetrized (with respect to interchange of mode), we compute the size of the symmetrized subspace and show that it too grows exponentially with the number of physical resources.  We remark that while a classical pachinko machine (or \lq\lq{}Galton’s board\rq\rq{}) will also have an exponential large state space, because no interference occurs there is only a quadratic increase with $L$ in the number of calculations necessary to simulate the output (corresponding to the number of beam splitters in the interferometer).   From this result we conclude that it is very likely that any classical computer that tries to simulate the operation of the quantum pachinko machine will always suffer an exponential slowdown. We will also show that no exponential growth occurs if Fock states are replaced with photonic coherent states or squeezed states, which elucidates part of the special nature of photonic Fock states. However an exponentially large Hilbert space, while necessary for classical non\textendash simulatability, is not sufficient. We then finally examine the physical symmetry requirements for bosonic versus fermionic multi\textendash particle states and show that in the bosonic case, in order to simulate the interferometer as a physics experiment, one must compute the permanent of a large matrix, which in turn is a problem in a computer algorithm complexity class strongly believed to be intractable on a classical or even a quantum computer. This concludes our elementary argument, which invokes only simple quantum mechanics, combinatorics, and a simplistic appeal to complexity theory.

\section{The Pachinko Machine Model}
As our argument is all about counting resources, we have carefully labeled all the components in the pachinko machine in Fig. \ref{fig:fig1} to help us with that reckoning. The machine has a total of $L$ levels of physical depth $d$ each. The input state at the top is the dual\textendash Fock state $\ket{N}^0_1\ket{M}^0_2$, where the superscripts label the level number and the subscripts the element in the row at that level (from left to right). We illustrate a machine of total numerical depth of $L = 3$. For $1\leq \ell < L$, we show the vacuum input modes along the edges of the machine. The resources we are most concerned about are energy, time, spatial dimension, and number of physical elements needed to construct the device. All of these scale either linearly or quadratically in either $L$ or $N + M$. The total physical depth is $D = L d$ and so the spatial area is $A = (\sqrt{2} D)^2=2L^2 d^2$. Using identical photons of frequency $\omega$, the energy per run is $E=(N+M)\hbar \omega$. The time it takes for the photons to arrive at the detectors is $T = \sqrt{2} L d / c$, where $c$ is the speed of light. In each level the photons encounter $\ell$ number of beam splitter (BS) so the total number is $ \# B = \sum_{\ell=1}^{L} \ell = L(L+1)/2$. Below each BS (with the exception of the $L$th level) there are two independently tunable phase shifters (PS) for a total number of PS that is $ \# \varphi = \sum_{\ell=1}^{L-1} 2\ell = L(L-1)$. The total number of detectors is $\# \nabla = 2L$. The total number of input modes is equal to the total number of output modes and is $\#I = \#O = 2L$. The total number of internal modes is $\# \psi = \sum_{\ell=1}^{L-1} 2\ell = L(L-1)$. As promised everything scales either linearly or quadratically in either $L$ or $N+M$.

The input state may be written in the Heisenberg picture as $\ket{N}^0_1\ket{M}^0_2= (\hdag{1}{0})^N (\hdag{2}{0})^M \ket{0}^0_1 \ket{0}^0_2/\sqrt{N!M!}$, where $\hat{a}^{\dagger}$ is a modal creation operator. Each BS performs a forward unitary mode transformation, which we illustrate with $B^1_1$, of the form $\hat{a}_1^1=i r_1^1 \hat{a}_1^0+ t_1^1 \hat{a}_2^0$ and $\hat{a}_2^1= t_1^1\hat{a}_1^0+i r_1^1\hat{a}_2^0$ where the reflection and transmission coefficients $r$ and $t$ are positive real numbers such that $ r^2 +t^2 = R+T =1$. The choice $r = t = 1/\sqrt{2}$ implements a 50\textendash 50 BS. Each PS is implemented by, for example, applying the unitary operation $\textrm{exp}(i \varphi_1^1 \hat{n}_1^1)$ on mode $\ket{\psi}_1^1$, where $\hat{n}_1^1 := \hdag{1}{1}\hat{a}_1^1$ is the number operator, $\hat{a}_1^1$ is the annihilation operator conjugate to $\hdag{1}{1}$, and $\varphi_1^1$ is a real number. Finally the $2L$ detectors in the final level $L$ are each photon number resolving \cite{lee}.

To argue that this machine (or any like it) cannot be simulated classically, in general, it suffices to show that this is so for a particular simplified example. We now take $N$ and $L$ arbitrary but $M = 0$ and turn off all the phase shifts and make all the BS identical by setting $ \varphi_k^{\ell}=0$, $t_k^{\ell} = t$, and $r_k^{\ell} =r$ for all $(k,\ell)$. We then need the backwards BS transformation on the creation operators, which is, $\hdag{1}{0} = i r \hdag{1}{1} + t \hdag{2}{1}$ and $\hdag{2}{0} = t \hdag{1}{1} + i r \hdag{2}{1} $. Similar transforms apply down the machine at each level. With $M = 0$ the input simplifies to $\ket{N}_1^0\ket{0}_2^0 = (\hdag{1}{0})^N \ket{0}_1^0\ket{0}_2^0/\sqrt{N!}$  and now we apply the first backwards BS transformation $\ket{\psi}_1^1\ket{\psi}_2^1= (i r \hdag{1}{1} + t\hdag{2}{1})^N\ket{0}_1^0\ket{0}_2^0/\sqrt{N!}$ to get the state at level one.

At every new level each $\hat{a}^\dagger$ will again bifurcate according to the BS transformations for that level, with the total number of bifurcations equal to the total number of BS, and so the computation of all the terms at the final level involves a polynomial number of steps in $L$. It is instructive to carry this process out explicitly to level $L = 3$ to get,
\begin{equation}
\begin{split}
\ket{\psi}^3&= \frac{1}{\sqrt{N!}}(i r t^2\hdag{1}{3} -r^2 t\hdag{2}{3}+i r (t^2-r^2)\hdag{3}{3} \\
&-2r^2 t \hdag{4}{3}+i r t^2 \hdag{5}{3} +t^3 \hdag{6}{3})^N \prod_{\ell=1}^6\ket{0}_{\ell}^3,
\end{split}
\label{eq:prod1}
\end{equation}
where we have used a tensor product notation for the states. If $r\cong0$ or $r\cong1$ the state is easily computed. Since we are seeking a regime that cannot be simulated classically we work with $r\cong t \cong 1/\sqrt{2}$.

\section{Solution in the Heisenberg and Schr\"{o}dinger Pictures}
It is now clear from Eq.(\ref{eq:prod1}) what the general form of the solution will be. We define
\begin{equation}
\ket{\psi}^L := \underset{\mathclap{\begin{subarray}{c}
\lbrace n_\ell\rbrace \\
N=\sum_{\ell=1}^{2L} n_\ell 
\end{subarray}}} 
 \sum \ket{\psi}_{\lbrace n_\ell\rbrace}^L \quad , \quad
\ket{0}^L := \prod_{\ell=1}^{2L}\ket{0}_{\ell}^L ,
\label{eq:definitions}
\end{equation}

and the general solution has the form,
\begin{equation}
\begin{split}
\ket{\psi}^L&=\frac{1}{\sqrt{N!}}\left( \sum_{\ell=1}^{2L}\alpha_{\ell}^L\hdag{\ell}{L} \right)^N \ket{0}^L \\
&=\frac{1}{\sqrt{N!}}\sum_{N=\sum_{\ell=1}^{2L} n_{\ell}} 
\binom{N}{n_1, n_2,\mathellipsis, n_{2L}} \\
&\times \prod_{1\leq k \leq 2L} (\alpha_{k}^L \hdag{k}{L})^{n_k}\ket{0}^L ,
\end{split}
\label{eq:state}
\end{equation}
where all the coefficients $\alpha_{\ell}^L$ will be nonzero in general. Since all the operators commute, as they each operate on a different mode, we have expanded Eq. (\ref{eq:state}) using the multinomial theorem where the sum in the expansion is over all combinations of non\textendash negative integers constrained by $N=\sum_{\ell=1}^{2L} n_l$ and
\begin{equation}
\binom{N}{n_1, n_2,\mathellipsis, n_{2L}}= \frac{N!}{n_1! n_2! \mathellipsis n_{2L}!}
\label{eq:binom}
\end{equation}
is the multinomial coefficient \cite{nist}. The state $\ket{\psi}^L$ is highly entangled over the number\textendash path degrees of freedom. Each monomial in the expansion of Eq. (\ref{eq:state}) is unique and so the action of the set of all monomial operators on the vacuum will produce a complete orthonormal basis set for the Hilbert space at level $L$, given by $ \ket{\psi}_{\lbrace n_\ell \rbrace}^L:=\prod_{\ell=1}^{2L}\ket{n_\ell}_{\ell}^L$, where the $n_\ell$ are subject to the same sum constraint. Let us call the dimension of that Hilbert space dim$[H(N,L)]$, which is therefore the total number of such basis vectors.

Taking $L=3$ and $N=2$, we can use Eq.(\ref{eq:state}) to compute the probability a particular sequence of detectors will fire with particular photon numbers. What is the probability detector one gets one photon, detector two also gets one, and all the rest get zero? This is the modulus squared of the probability amplitude of the state $\ket{1}_1^3\ket{1}_2^3\ket{0}_3^3\ket{0}_4^3\ket{0}_5^3\ket{0}_6^3$. Setting $r=t=1/\sqrt{2}$ for the 50\textendash50 BS case, from Eq.(\ref{eq:prod1}) we read off $\alpha_1^3=irt^2=i/(2\sqrt{2})$ and $\alpha_2^3=-r^2t=-1/(2\sqrt{2})$, and so the probability of this event is given by $P_{110000}\cong 0.031$.

It turns out that it is possible (for general $L$ and $N$) to compute the single and binary joint probabilities, that detector $p$ gets $n$ photons and detector $q$ gets $m$ \cite{mayer}. However computing arbitrary joint probabilities between triplets, quadruplets, etc., of detectors rapidly becomes intractable. We can provide a closed form expression for dim$[H(N,L)]$ by realizing that it is the same as the number of different ways one can add up non\textendash negative integers that total to fixed $N$. More physically this is the number of possible ways that $N$ indistinguishable photons may be distributed over $2L$ detectors. The answer is well known in the theory of combinatorics and is:
\begin{equation}
\textrm{dim}[H(N,L)]=\binom{N+2L-1}{N},
\label{eq:dim1}
\end{equation}
where this is the ordinary binomial coefficient \cite{benjamin}. For our example with $L = 3$, $N = 2$, Eq.(\ref{eq:dim1}) implies that the number of distinct probabilities $P_{npqrst}^3$ to be tabulated is again 21.

We first examine two \lq\lq{}computationally simple\rq\rq{} examples. Taking $N$ arbitrary and $L =1$ we get dim$[H(N,1)]=N+1$, which is easily seen to be the number of ways to distribute $N$ photons over two detectors. Next taking $N = 1$ and $L$ arbitrary we get dim$[H(1,L)]=2L$, which is the number of ways to distribute a single photon over $2L$ detectors. If we were to invoke Dirac's edict\textemdash \lq\lq{}Each photon then interferes only with itself.\rq\rq{}\textemdash we would then expect that adding a second photon should only double this latter result \cite{dirac1}. Instead the effect of two\textendash photon interference on the state space can be seen immediately by computing dim$[H(2,L)]=L(2L+1)$. That is, adding a second photon causes a quadratic (as opposed to linear) jump in the size of the Hilbert space. Dirac was wrong; photons do interfere with each other, and that multiphoton interference directly affects the computational complexity. All these three cases are simulatable in polynomial time steps with $N$ and $L$, but we see a quadratic jump in dimension as soon as we go from one to two photons. These jumps in complexity continue for each additional photon added and the dimension grows rapidly.

We therefore next investigate a \lq\lq{}computationally complex\rq\rq{} intermediate regime by fixing $N=2L-1$. That is we build a machine with total number of levels $L$ and then choose an odd\textendash numbered photon input so that this restriction holds. Equation (\ref{eq:dim1}) becomes dim$[H(N)]=(2N)!/(N!)^2$. Deploying Sterling's approximation for large $N$, in the form $n! \cong (n/e)^n \sqrt{2 \pi n}$  we have dim$[H(N)]\cong2^{2N}/\sqrt{\pi N}$. This is one of our primary results. The Hilbert space dimension scales exponentially with $N=2L-1$. Since all the physical parameters needed to construct and run our quantum pachinko machine scale only linear or quadratically with respect to $N$ or $L$, we have an exponentially large Hilbert space produced from a polynomial number of physical resources \textemdash Feynman's necessary condition for a potential universal quantum computer.

Let us suppose we build onto an integrated optical circuit a machine of depth $L = 69$ and fix $N = 2L-1 = 137$. Such a device is not too far off on the current quantum optical technological growth curve \cite{bonneau}. Then we have dim$[H(137)]=10^{81}$, which is again on the order of the number of atoms in the observable universe. Following Feynman's lead, we conclude that, due to this exponentially large Hilbert space, we have a sufficient condition that a classical computer can not likely efficiently simulate this device. However this is not a necessary condition. From the Gottesman-Knill theorem we know that quantum circuits that access an exponentially large Hilbert space may sometimes be efficiently simulated \cite{gottesman}. We will strengthen our argument (below) by discussing the necessity of properly symmetrizing a multi-particle bosonic state and tie that physical observation back to the complexity of computing the permanent of a large matrix.

Let us now compare our Heisenberg picture result to that of the Schr\"{o}dinger picture. In the computationally complex regime where $N=2L-1$ the number of distinct Feynman\textendash like paths we must follow in the Schr\"{o}dinger picture is $2^{LN}=2^{N(N+1)/2}\cong 2^{N^2/2}$. Taking $N = 137$ and $L = 69$, as in the previous example, we get an astounding $2^{9453}\cong 4\times10^{2845}$ total paths. Dirac proved that the Heisenberg and Schr\"{o}dinger pictures are mathematically equivalent, that they always give the same predictions, but we see here that they are not always necessarily \textit{computationally} equivalent \cite{dirac2}. Calculations in the Heisenberg picture are often \textit{much} simpler than in the Schr\"{o}dinger picture. The fact that the two pictures are not always computationally equivalent is implicit in the Gottesman\textendash Knill theorem; however, it is satisfying to see here just how that is so in a simple optical interferometer \cite{gottesman}.

\section{Sampling with Coherent \& Squeezed State Inputs}
To contrast this exponential overhead from the resource of bosonic Fock states, let us now carry out the same analysis with the bosonic coherent input state input $\ket{\beta}_1^0\ket{0}_2^0$, where we take the mean number of photons to be $\abs{\beta}^2=\overline{n}$. In the Heisenberg picture this input becomes $\hat{D}_1^0(\beta)\ket{0}_1^0\ket{0}_2^0$, where $\hat{D}_1^0(\beta)=\textrm{exp}(\beta \hdag{1}{0}-\beta^* \hat{a}_1^0)$ is the displacement operator \cite{gerry}. Applying the BS transformations down to final level $L$ we get
\begin{equation}
\begin{split}
\ket{\psi}^L&=\textrm{exp}\left(\beta \sum_{\ell=1}^{2L}\alpha_{\ell}^L \hdag{\ell}{L}- \beta^* \sum_{\ell=1}^{2L}\alpha_{\ell}^{L*} \hat{a}_{\ell}^L\right)\ket{0}^L \\
&=\prod_{\ell=1}^{2L}\textrm{exp}(\beta \alpha_{\ell}^L \hdag{\ell}{L}- \beta^*\alpha_{\ell}^{L*}\hat{a}_{\ell}^L)\ket{0}^L \\
&=\prod_{\ell=1}^{2L}\ket{\beta \alpha_{\ell}^L}_{\ell}^L .
\end{split}
\label{eq:state2}
\end{equation}
At the output we have $2L$ coherent states that have been modified in phase and amplitude. This is to be expected, as it is well known that linear interferometers transform a coherent state into another coherent state. Since all the coefficients $\alpha_{\ell}^L$ are computable in $\#B=L(L+1)/2$ steps, this result is obtained in polynomial time steps in $L$, independent of $\overline{n}$. The mean number of photons at each detector is then simply $\overline{n}_{\ell}^L=\abs{\beta \alpha_{\ell}^L}^2=\overline{n}\abs{\alpha_{\ell}^L}^2$.

A similar analysis may be carried out for bosonic squeezed input states. Taking, for example, a single\textendash mode squeezed vacuum input $\ket{\xi}_1^0\ket{0}_2^0= \hat{S}_1^0(\xi)\ket{0}_1^0\ket{0}_2^0$, with the squeezing operator defined as $\hat{S}_1^0(\xi)=\textrm{exp}\lbrace [\xi^*(\hat{a}_1^0)^2-\xi(\hdag{1}{0})^2]/2\rbrace$, we arrive at,
\begin{equation}
\ket{\psi}^L=\textrm{exp}\left\{ \left[\xi^*\left(\sum_{\ell=1}^{2L}\alpha_{\ell}^*\hat{a}_{\ell}^L\right)^2-\xi \left(\sum_{\ell=1}^{2L}\alpha_{\ell}\hdag{\ell}{L}\right)^2\right]/2\right\} \ket{0}^L,
\label{eq:state3}
\end{equation}
which does not in general decompose into a separable product of single\textendash mode squeezers on each output port. Nevertheless the probability amplitudes may still be computed in a time polynomial in $L$ by noting that, from Eq.(\ref{eq:dim1}) with $N = 2$, there are at most $2L(L+1)$ terms in this exponent that must be evaluated. This result generalizes to arbitrary Gaussian state inputs \cite{bartlett,*veicht}. The output of the interferometer may be then calculated on the transformed device in polynomial steps in $L$.

The exponential scaling comes from the bosonic Fock structure $\ket{N}=(\hat{a}^\dagger)^N\ket{0}/\sqrt{N!}$ and the rapid growth of the number\textendash path entanglement in the interferometer. It is well known that beam splitters can generate number\textendash path entanglement from separable bosonic Fock states. For example, the simplest version of the HOM effect at level one with separable input $\ket{1}_1^0\ket{1}_2^0$ becomes $\ket{\psi}_1^1 \ket{\psi}_2^1=(i\hdag{1}{1}+\hdag{2}{1})(\hdag{1}{1}+i\hdag{2}{1})\ket{0}_1^1\ket{0}_2^1/2=i[\ket{2}_1^1\ket{0}_2^1+\ket{0}_1^1\ket{2}_2^1]/\sqrt{2}$ a NOON state \cite{ou}. Such entangled NOON states violate a Bell inequality and are hence nonlocal even though the input was not \cite{wildfeuer}. For arbitrary bosonic Fock input states and interferometer size the amount of number\textendash path entanglement grows exponentially fast. However, even in the case of fermionic interferometers, where there is a restriction of two identical particles per mode, the Hilbert space can still grow exponentially fast (just not quite as fast as in the case of bosons) as we shall now show. 

\section{Comparison of Bosonic to Fermionic Fock State Inputs}
We now compare the multimode bosonic Fock state interferometer to the multimode fermionic interferometer. We will restrict ourselves to spin\textendash1/2 neutral fermions such as neutrons that are commonly used in interferometry. Now the number of fermions per input mode is restricted to zero, one, or two and we can have two only if they have opposite spin states to be consistent with the Pauli exclusion principle. The exclusion principle is derived from the requirement that the total multi\textendash particle fermionic wave function, which is the product of the spin and spatial wave functions, is antisymmetric under the interchange of any two particle state labels. Likewise there is a constraint on the bosonic multi\textendash particle multi\textendash mode wave function that the total wave function be symmetric. The symmetry of the wave function must be enforced at each beam splitter where the particles become indistinguishable and the spatial part of the wave function experiences maximal overlap for multi\textendash particle interference to occur. For the sake of argument we take the coherence length of the particles to be infinite (or at least much larger than the depth of the interferometer $Ld$) so that enforcing the correct symmetry at each beam splitter requires enforcing the correct symmetry everywhere in space.

Some care must now be used in the notation. For example when we write the bosonic spatial wave function input state $\ket{1}^{b}_{A_{\text{in}}} \ket{1}^b_{B_{\text{in}}}$, we are assuming both bosons have the same spin state, since clearly this state is spatially symmetric under particle interchange its spin state must also be, so that the product of the two (total wave function) remains symmetric. To denote this point we instead write $\ket{\uparrow}^{b}_{A_{\text{in}}} \ket{\uparrow}^b_{B_{\text{in}}}$ to explicitly show the spin state. [More properly we should write $\psi^b(x_{A_{\text{in}}})\psi^b(x_{B_{\text{in}}})\ket{\uparrow}^{b}_{A_{\text{in}}} \ket{\uparrow}^b_{B_{\text{in}}}$  but this notation is a bit cumbersome.] Thence for a 50:50 BS the HOM effect for bosons in the same spin state can be written, $\ket{\uparrow}^{b}_{A_{\text{in}}} \ket{\uparrow}^b_{B_{\text{in}}}\overset{BS}{\rightarrow}  \ket{\uparrow \uparrow}^b_{A_{\text{out}}}\ket{0}^b_{B_{\text{out}}} + \ket{0}^b_{A_{\text{out}}}\ket{\uparrow \uparrow}^b_{B_{\text{out}}}$ , so both bosons \lq\lq{}stick\rq\rq{} at the beam splitter and emerge together. This effect arises as a direct result of the fact that the \textit{spatial} part of the wave function, which gives rises to an effective attraction at the BS, is symmetric. We could instead prepare an antisymmetric bosonic singlet spin state input $\ket{\uparrow}^{b}_{A_{\text{in}}} \ket{\downarrow}^b_{B_{\text{in}}}- \ket{\downarrow}^{b}_{A_{\text{in}}} \ket{\uparrow}^b_{B_{\text{in}}}$, in which case the spatial wavefunction must be also antisymmetric, $\psi^b(x_{A_{\text{in}}})\psi^b(x_{B_{\text{in}}})-\psi^b(x_{B_{\text{in}}})\psi^b(x_{A_{\text{in}}})$ , so that the product of the two remains symmetric. In this case the particles behave fermionically as far as the spatial wavefunction overlap is concerned at the BS and they repel each other in an anti\textendash HOM effect, always exiting out separate ports and never together;  $\ket{\uparrow}^{b}_{A_{\text{in}}} \ket{\downarrow}^b_{B_{\text{in}}}- \ket{\downarrow}^{b}_{A_{\text{in}}} \ket{\uparrow}^b_{B_{\text{in}}} \overset{BS}{\rightarrow} \ket{\uparrow}^{b}_{A_{\text{out}}} \ket{\downarrow}^b_{B_{\text{out}}}- \ket{\downarrow}^{b}_{A_{\text{out}}} \ket{\uparrow}^b_{B_{\text{out}}}$    \cite{loudon1,loudon2}. The reverse happens for fermions.  

For example, the symmetric spin input state $\ket{\uparrow}^{f}_{A_{\text{in}}} \ket{\uparrow}^f_{B_{\text{in}}}$  is allowed for fermions only if the spatial wave wave function is antisymmetric, $\psi^f(x_{A_{\text{in}}})\psi^f(x_{B_{\text{in}}})-\psi^f(x_{B_{\text{in}}})\psi^f(x_{A_{\text{in}}})$, so that the entire wave function product remains antisymmetric. Since the spatial part governs the HOM effect they repel at the BS and obey an anti\textendash HOM effect and always exit out separate ports, consistent with the exclusion principle, namely $\ket{\uparrow}^{f}_{A_{\text{in}}} \ket{\uparrow}^f_{B_{\text{in}}} \overset{BS}{\rightarrow}\ket{\uparrow}^{f}_{A_{\text{out}}} \ket{\uparrow}^f_{B_{\text{out}}}$. However, we can make the fermions behave spatially bosonically by preparing them in a spin\textendash antisymmetric singlet input state, which then must be symmetric in the spatial part, and so they behave as bosons as far as the spatial overlap is concerned, and we recover the usual HOM effect, where now they always exit the same port together: $\ket{\uparrow}^{f}_{A_{\text{in}}} \ket{\downarrow}^f_{B_{\text{in}}}- \ket{\downarrow}^{f}_{A_{\text{in}}} \ket{\uparrow}^f_{B_{\text{in}}} \rightarrow \ket{\uparrow \downarrow}^{f}_{A_{\text{out}}} \ket{0}^f_{B_{\text{out}}}- \ket{\downarrow \uparrow}^{f}_{A_{\text{out}}} \ket{0}^f_{B_{\text{out}}}$. There is no violation of the exclusion principle as they also always exit with opposite spins. (This type of effective spatial attraction between fermions in a spin singlet state explains why the ground state of the neutral hydrogen molecule is a bound state.) It is clear then that even fermions can experience number-path entanglement in a linear interferometer, although not to the same degree as bosons. However this entanglement is still sufficient to lead to an exponential growth in the fermionic Hilbert space, as we shall now argue.

Now we are ready to apply our resource counting argument to the fermionic case. For fermions the computationally complex regime may be accessed when the number of input particles $N$ is half the number of input modes $2L$. The dimension of the Hilbert space may be computed as before and turns out to be, for this example, $\binom{2L}{N}$. This also grows exponentially as a function of the resources choosing $N = L$. Following the same Sterling's approximation argument as above we get exactly the same exponential formula for the Hilbert space dimension as with bosons, namely $ 2^{2N}/ \sqrt{\pi N}$.

So, in general, in both the fermionic and bosonic case the Hilbert space dimension grows exponentially with respect to the resources: particle number and mode number. However, Feynman's arguments notwithstanding, an exponential growth in the Hilbert space is only sufficient but not necessary to attain classical non\textendash simulatability. For example, from the Gottesman\textendash Knill theorem, we can construct a Clifford\textendash algebra\textendash based quantum computer circuit that accesses an exponentially large Hilbert space but still can be simulated efficiently classically \cite{gottesman}. Sometimes there are shortcuts through Hilbert space, as we shall now argue is the case here for fermions but not for bosons. 

In order to access these large Hilbert spaces in the interferometer one must require that multi\textendash particle interference take place at each beam splitter, where the particles must be indistinguishable, and the spatial wave function overlap determines the type of particle\textendash mode entanglement that will result. The overall bosonic wave function (spatial multiplied by spin) must be totally symmetric and the overall fermionic wave function must be totally antisymmetric at each row of BS, and so they must have these symmetries everywhere in space and particularly at the input. Now if we give up on a complete tabulation of the Hilbert state space at level $L$, due to its exponential growth, and treat the interferometer using a standard quantum optical input\textendash output formalism, there is an efficient way to take a given multi\textendash particle, multi\textendash mode input state at the top of the interferometer to the bottom of the interferometer. This method is called matrix transfer and is accomplished by encoding each level of BS transformations in terms of $L$ matrices of size $(2L) \times (2L)$ and then multiplying them together. This can be done in the order of $O(L^3)$ steps and so it is efficient.

We must now address the issue of computing the sampling output of the interferometer.  While the one- and two- particle joint particle detection probabilities at the detectors may be computed efficiently, computation of the higher order joint probabilities rapidly become intractable \cite{mayer}.  In order to compute the complete joint probability distribution, we must compute the determinant (if the input is fermionic) or the permanent (if the input is bosonic) of the $(2L) \times (2L)$ matrix found above. Using the method of Laplace decomposition for constructing the determinant of a matrix, one decomposes the large determinant into a sum over ever-smaller determinants, appending alternating plus and minus signs to each in a checkerboard pattern. Constructing the permanent follows the same process but
all the signs are now only plus.

However, for the determinant, there is a polynomial shortcut through the exponential Hilbert space\textemdash the row\textendash reduction method. For fermions we may always construct the most general input state efficiently.  On the other hand, there is no known method such as row reduction to compute the permanent of an arbitrary matrix efficiently. The most efficient known protocols for the permanent computation are variants on the Laplace decomposition and all scale exponentially with the size of the matrix. This problem of computing the permanent, in the lingo of computer complexity theory, is that it is in the class of \lq\lq{}\#P\textendash hard\rq\rq{} (sharp\textendash P\textendash hard) problems. All problems in this class are very strongly believed to be intractable on any classical computer and also strongly suspected to be intractable on even a quantum computer \cite{aaronson}. While some matrices have a general form for which the permanent can be more easily computed, for an arbitrary interferometer setup, this matrix does not have a general form which we can exploit in order to shortcut the computation of the permanent.  We are left with the task of using our most efficient, general, exact permanent computing algorithm (Ryser's formula), which requires $O(2^{2L}L^2)$ number of steps \cite{ryser}. Finally, we have reached the snag that undermines our ability to efficiently compute the output and so renders simulation of the device classically intractable.

\section{Conclusion}
In conclusion, we have shown that a multi\textendash mode linear optical interferometer with arbitrary Fock input states is very likely not simulatable classically. Our result is consistent with the argument of AA. Without invoking much complexity theory, we have argued this by explicitly constructing the Hilbert state space of a particular such interferometer and showed that the dimension grows exponentially with the size of the machine. The output state is highly entangled in the photon number and path degrees of freedom. We have also shown that simulating the device has radically different computational overheads in the Heisenberg versus the Schr\"{o}dinger picture, illustrating just how the two pictures are not in general computationally equivalent within this simple linear optical example. Finally we supplement our Hilbert space dimension argument with a discussion of the symmetry requirements of multi\textendash particle interferometers and particularly tie the simulation of the bosonic device to the computation of the permanent of a large matrix, which is strongly believed to be intractable. It is unknown (but thought unlikely) if such bosonic multi\textendash mode interferometers as these are universal quantum computers, but regardless they will certainly not be fault tolerant. As pointed out by Rohde \cite{rohde}, it is well known that Fock states of high photon number are particularly sensitive to loss \cite{huver}. They are also super\textendash sensitive to dephasing as well \cite{qasimi}. This implies that even if such a machine turns out to be universal it would require some type of error correction to run fault tolerantly. Nevertheless, such devices could be interesting tools for studying the relationship between multi\textendash photon interference and quantum information processing for small numbers of photons. If we choose each of the PS and BS transformations independent of each other, we have a mechanism to program the pachinko machine by steering the output into any of the possible output states. Even if universality turns out to be lacking we may very well be able to exploit this programmability to make a special purpose quantum simulator for certain physics problems such as frustrated spin systems \cite{britton}.  
\begin{acknowledgments}
B.T.G would like to acknowledge support from the NPSC/NIST fellowship program. J.P.D. would like to acknowledge support from the NSF. This work is also supported by the Intelligence Advanced Research Projects Activity (IARPA) via Department of Interior National Business Center contract number D12PC00527. The U.S. Government is authorized to reproduce and distribute reprints for governmental purposes notwithstanding any copyright annotation thereon. Disclaimer: The views and conclusions contained herein are those of the authors and should not be interpreted as necessarily representing the official policies or endorsements, either expressed or implied, of IARPA, DoI/NBC, or the U.S. Government. We also would like to acknowledge interesting and useful discussions with S. Aaronson, S. T. Flammia, K. R. Motes, and P. P. Rohde. 
\end{acknowledgments}

\bibliography{Gard13tst}

\end{document}